\documentclass{article}
\usepackage{ijcai22}

\usepackage{times}
\usepackage{soul}
\usepackage{url}
\usepackage[hidelinks]{hyperref}
\usepackage[utf8]{inputenc}
\usepackage[small]{caption}
\usepackage{graphicx}
\usepackage{amsmath}
\usepackage{amsthm}
\usepackage{booktabs}
\usepackage{algorithm}
\usepackage{algorithmic}

\usepackage{amssymb} % for fix Undefined control sequence by tao

\usepackage{amsmath} % for fix environment align undefined by tao

\usepackage{caption} % for sub-figure by tao
\usepackage{subcaption} % for sub-figure by tao

\title{DAGAM: A Domain Adversarial Graph Attention Model for Subject Independent EEG-Based Emotion Recognition}

\author{
Tao Xu$^1$
\and
Wang Dang$^1$\and
Jiabao Wang$^1$\And
Yun Zhou$^2$
\affiliations
$^1$School of Software, Northwestern Polytechnic University\\
$^2$Faculty of Education, Shaanxi Normal University\\
\emails
xutao@nwpu.edu.cn,
{dwnpu,wjiabao}@mail.nwpu.edu.cn,
zhouyun@snnu.edu.cn
}
\begin{document}

\maketitle

\begin{abstract}
% One of the most significant challenges of EEG-based emotion recognition is the cross-subject EEG variations, leading to poor performance and generalizability. In this paper, we propose a novel EEG-based emotion recognition model called domain adversarial graph attention model (DAGAM). The basic idea is to generate a graph to model multichannel EEG signals using biological topology. Graph theory can topologically describe and analyze relationships and mutual dependency between channels of EEG. Then, unlike other graph convolutional networks, self-attention pooling is applied to benefit salient EEG feature extraction from the graph, which effectively improves the performance. Finally, after graph pooling, the domain adversarial based on the graph is employed to identify and handle EEG variation across subjects, efficiently reaching good generalizability. We conduct extensive evaluations on two benchmark datasets (SEED and SEED IV) and obtain state-of-the-art results in subject-independent emotion recognition. Our model boosts the SEED accuracy to 92.59\% (4.69\% improvement) with the lowest standard deviation of 3.21\% (2.92\% decrements), and SEED IV accuracy to 80.74\% (6.90\% improvement) with the lowest standard deviation of 4.14\% (3.88\% decrements) respectively.

One of the most significant challenges of EEG-based emotion recognition is the cross-subject EEG variations, leading to poor performance and generalizability. This paper proposes a novel EEG-based emotion recognition model called the domain adversarial graph attention model (DAGAM). The basic idea is to generate a graph to model multichannel EEG signals using biological topology. Graph theory can topologically describe and analyze relationships and mutual dependency between channels of EEG. Then, unlike other graph convolutional networks, self-attention pooling is applied to benefit salient EEG feature extraction from the graph, which effectively improves the performance. Finally, after graph pooling, the domain adversarial based on the graph is employed to identify and handle EEG variation across subjects, efficiently reaching good generalizability. We conduct extensive evaluations on two benchmark datasets (SEED and SEED IV) and obtain state-of-the-art results in subject-independent emotion recognition. Our model boosts the SEED accuracy to 92.59\% (4.69\% improvement) with the lowest standard deviation of 3.21\% (2.92\% decrements) and SEED IV accuracy to 80.74\% (6.90\% improvement) with the lowest standard deviation of 4.14\% (3.88\% decrements) respectively.

\end{abstract}

\section{Introduction}

\noindent Investigating emotion recognition is a continuing concern within computer science. The findings and products of this emerging focus are increasingly applied to education, digital games, e-commerce, ad, e-health, and many other areas. Electroencephalogram (EEG) has been suggested as a promising tool to investigate human emotions as it can directly and precisely reflect cognitive and emotional states with relatively low costs. Thus, EEG-based emotion recognition has attracted considerable research attention and interest.

However, studies of applying deep learning algorithms to EEG-based subject-independent emotion recognition are unsatisfactory. First, multichannel EEG signals have a structure based on biological topography, belonging to a non-Euclidean domain. Directly applying deep learning methods to EEG-based Recognition does not work well since these methods are designed for the tasks of CV and NLP. Second, the EEG signals vary significantly between individuals, leading to the different distribution of source domain and target domain. This makes it challenging to achieve good performance across subjects.

The human brain's structural and functional systems have features of biological topography. Graph theory can topologically describe and analyze relationships and mutual dependency between channels of EEG. The Graph Neural Networks (GNN)
\cite{scarselli_graph_2009} make it promising to solve the classification problems on EEG data. Based on graph, many researchers have made great efforts \cite{song_eeg_2020} \cite{zhong_eeg-based_2020}. These graph-based methods try to learn and extract the most salient features from the whole high dimensional graph feature space generated by EEG data. Although some existing research starts to recognize the critical role of EEG channels' topology, they do not fully utilize such structure to learn salient EEG features well.

When EEG training and testing data from different individuals, most current recognition methods did not perform well. For EEG-based emotion recognition, the source domain and target domains' data distribution is different. This issue can be considered as a kind of domain adaptation. Ganin et al. \cite{ganin_domain-adversarial_2016} proposed a domain-adversarial training of neural networks (DANN) to solve the cross-subject classification problem. Inspired by the idea of DANN, many studies have attempted and made achievements \cite{baoTwoLevelDomainAdaptation2021,zhong_eeg-based_2020,luoWGANDomainAdaptation2018}. However, there is still considerable room to improve the performance.

To address the two aforementioned issues on subject-independent emotion recognition, we propose a novel EEG-based emotion recognition model called domain adversarial graph attention model (DAGAM). First, we use a graph to model EEG signals based on biological topology. Then, the graph convolutional networks with self-attention pooling are applied to extract EEG features strongly correlated with emotions. Finally, after graph pooling, the domain adversarial based on the graph is employed to identify emotions across subjects. The main contributions lie in the following aspects:
\begin{itemize}
\item The basic idea of DAGAM is to generate a graph to model multichannel EEG signals using biological topology. The use of graph attention neural networks (GANN) effectively explores the relationships among multiple EEG channels for emotion recognition. Unlike other graph convolutional networks, self-attention pooling is applied to benefit salient EEG feature extraction from the graph, which effectively improves the performance.
\item The domain adversarial (DA) based on the graph is employed to identify and handle EEG variation across subjects. Combining DA and GANN, the source domain and the target domain can adapt to each other.
\end{itemize}

After evaluating DAGAM on two public emotion EEG datasets: SEED \cite{zheng_investigating_2015}, SEED IV \cite{zheng_emotionmeter_2019}, we found that our model has achieved the state of the art results in subject-independent emotion recognition, reaching a superior accuracy of performance with the lowest standard deviation (SEED: 92.59\%/3.21\%, SEED-IV: 80.74\%/4.14\%) compared to other methods.

\section{Related Work}
EEG-based emotion recognition has received increased attention in recent years. The methods can be categorized into two groups. One group focuses on finding crucial features. Shi et al. \cite{shi_differential_2013} proposed a novel feature called differential entropy for EEG-based vigilance estimation. Jenke et al. \cite{jenke_feature_2014} reviewed a wide range of features to attempt to find suitable features for relevant emotions. Wang et al. \cite{wang_emotional_2014} compared three existing EEG features: power spectrum, wavelet, and nonlinear dynamical analysis for improving emotion recognition. Another group is committed to proposing better classification algorithms. Petrantonakis et al. proposed a robust emotion recognition method based on higher order crossings (HOC) analysis.
Zhang et al. \cite{zhang_variational_2020} proposed a heuristic Variational Pathway Reasoning (VPR) method to deal with EEG-based emotion recognition. Xu et al. \cite{DecodeBrainSystemADynamicAdaptiveConvolutionalQuorumVotingApproachforVariableLengthEEGData2020} proposed a dynamic adaptive convolutional quorum voting approach for variable-length EEG data.

Recently, increased researchers have paid attention to subject-independent emotion recognition. For example, Li et al. \cite{li_multisource_2020} proposed a multisource transfer learning method for cross-subject EEG emotion recognition. Li et al. \cite{liBihemisphereDomainAdversarial2018} proposed a bi-hemisphere domain adversarial neural network (BiDANN) model, which achieves good performance on cross-subject recognition. However, compared with the performance of subject-dependent emotion recognition, there is still room to improve for subject-independent emotion recognition.

After the first Graph Neural Network (GNN) was proposed in 2009 \cite{scarselli_graph_2009}, different GNNs are applied to different fields. EEG data are considered to belong to non-Euclidean domains, which can be represented as a graph. Graph model contains rich relational information \cite{zhou_graph_2020}, and can reflect the connections between different regions of the brain. At present, many researchers attempt to apply it to the domain of EEG-based emotion recognition. Song et al. \cite{song_eeg_2020} proposed novel dynamical graph convolutional neural networks. Zhong et al. \cite{zhong_eeg-based_2020} proposed a regularized graph neural network (RGNN) for EEG-based emotion recognition, which extracts both local and global features among different EEG channels based on the biological topology among different brain regions. These methods only benefit some known biological features, but did not try to use unknown crucial internal connections and features via learning. The direction of future research is how to graph model and attention mechanism to find the salient features of EEG signals related to emotions.

Domain-Adversarial Neural Network \cite{ganin_domain-adversarial_2016} was the first work to demonstrate the success for two distinctive classification problems in 2016. Soon, it was widely used in the field of emotion recognition. Li et al. \cite{liCrossSubjectEmotionRecognition2018} applied Deep Adaptation Network (DAN) to eliminate the individual differences in EEG signals. Luo et al. \cite{luoWGANDomainAdaptation2018} proposed a novel Wasserstein Generative Adversarial Network Domain Adaptation (WGANDA) framework for building cross-subject EEG-based emotion recognition models. Bao et al. \cite{baoTwoLevelDomainAdaptation2021} proposed a Two-level Domain Adaptation Neural Network (TDANN) to construct a transfer model for EEG-based emotion recognition. Zhao et al. \cite{zhao_plug-and-play_2021} proposed a plug-and-play domain adaptation method for dealing with the individual difference. It points out the research direction for subject-independent emotion recognition and finds a suitable classifier for domain adversarial.

\section{Domain Adversarial Graph Attention Model}
\begin{figure*}[!h]
\centering
\includegraphics[width=0.88\textwidth]{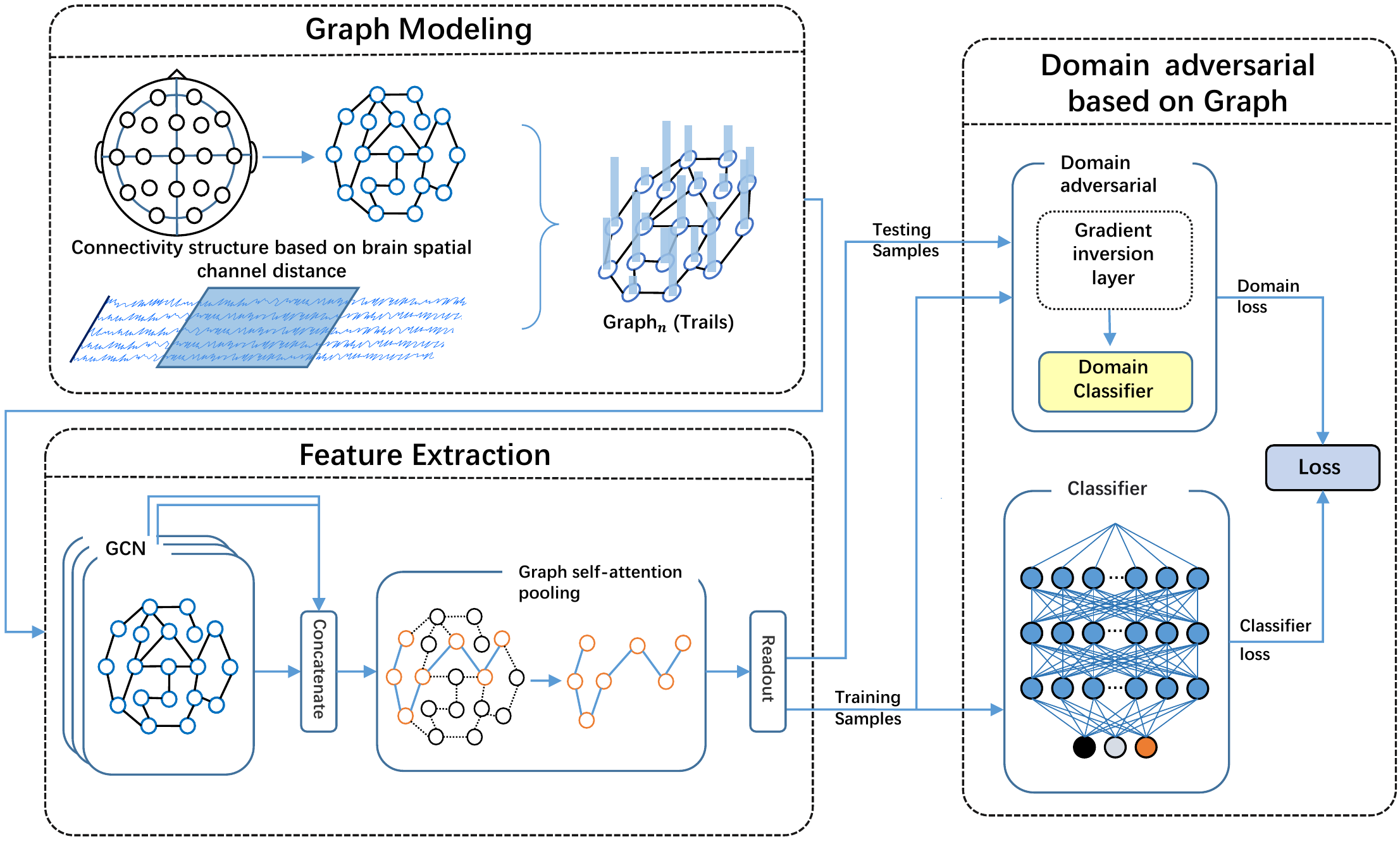} % Reduce the figure size so that it is slightly narrower than the column. Don't use precise values for figure width.This setup will avoid overfull boxes.
\caption{The structure of Domain Adversarial Graph Attention Model}.
\label{Fig:1}
\end{figure*}

The structure of Domain Adversarial Graph Attention Model (DAGAM) is as shown in Fig.\ref{Fig:1}. It contains three main parts: EEG data modeling based on the graph, graph attention neural networks, and domain adversarial based on the graph. First, the EEG data are modeled based on EEG channels' dependencies. Next, the graph attention neural is proposed to extract the core features and discard the unimportant channels in the graph structure. Finally, domain adversarial based on the graph helps to handle cross-subject EEG variations, enabling to achieve good performance on subject-independent emotion recognition. The detail of each part is provided as follows.

\subsection{EEG modeling based on the graph}
EEG data are collected by EEG cap based on the 10-20 system. Each electrode position represents the surface division of the brain, which interconnects and influences each other. The traditional Convolution Neural Networks (CNNs) can not directly benefit this biological topology. Graph Neural Networks provide an opportunity to tackle this issue.

The first thing is to model EEG signals by the graph. The basic graph can be expressed as a set of vertices and edges, denoted as $G=(V,E)$, where $V$ is the set of vertices and $E$ is the set of edges. In our work, the vertices set can be expressed as the matrix $X \in \mathbb{R}^{N \times D}$, and the edge set can be expressed as the adjacency matrix $A \in \mathbb{R}^{N \times D}$, where $N$ represents the number of EEG channels, $D$ represents EEG data over time.

$A$ is used to reflect the biological topography of EEG and indicates the relationship of EEG channels. The spatial distance between the channels is used to define the weight of the edge. To correctly reflect this kind of relationship, we attempt to define the element of the adjacency matrix based on the method proposed by \cite{zhong_eeg-based_2020}, as follows:

\begin{equation} \label{eq:1}
A_{ij} = min(1,\frac{\sigma}{d_{ij}^2})
\end{equation} where $d_{ij}$ notes the physical distance between channels $i$ and $j$, $\sigma$ is a constant, used to calibrate the weight $A_{ij}$ can fall within $(0,1)$

Several global connections are added to the adjacency matrix to improve network efficiency. The global connection depends on the specific electrode position used in the experiment. Previous biological studies have proved that the asymmetry of neuronal activity between the left and right hemispheres is informative in terms of potency and arousal prediction, which have supported the selection of global channels. To use this information, we initialize the global inter-channel relationship in A to $[-1,0]$.

% \begin{equation}\label{eq:pabc}
% A_{ij}=\left\{
% \begin{array}{rcl}
% w_{ij}    && {(v_i,v_j)\subseteq E}\\
% 0 && {(v_i,v_j)\not\subset E}\\
% \end{array} \right.
% \end{equation}

\subsection{Graph attention neural networks}
Graph structure helps us to model EEG data. However, long-period EEG data with a complicated graph structure brings high computational cost for emotion recognition. The pooling method considers features both in the channels and the whole structure of the graph and removes the influence of unimportant nodes. It attempts to use a reasonable number of parameters to obtain better graph classification performance. Inspired by the work of \cite{lee_self-attention_2019}, we adopted a graph pooling method based on self-attention, called SAGPool, to extract crucial features from EEG data. The detailed steps will be shown as follows:

First, the EEG data is processed by three layers of GNN to obtain the self-attention score. A widely used GNN model, Graph Convolutional Networks (GCN) \cite{kipf_semi-supervised_2017}, is implemented here, which is formulated as follows:
% \begin{equation}\label{eq:3}
% Y^{l+1}=GCN(Y^l,A)
% \end{equation}

\begin{equation}\label{eq:pabc}
Score =\sigma (\tilde{L}_{sym}XW_{att})
\end{equation} where $\sigma$ is the activation function of the layer network, and $W_{att}$ is a weighted matrix used to perform affine transformation on the input graph signal. $\tilde{L}_{sym}$ is the re-normalized Laplacian matrix following by \cite{kipf_semi-supervised_2017}.

The $index$ and $Score_{mask}$ of self-attention graph pooling can be obtained as follows:

% \begin{equation}\label{eq:pabc}
% index=top-rank(Score,[kN])
% \end{equation}
% \begin{equation}\label{eq:p}
% Score_{mask}=Score_{index}
% \end{equation}

\begin{equation}\label{eq:pabc}
\begin{aligned}
index=top-rank(Score,[kN])\\
Score_{mask}=Score_{index}
\end{aligned}
\end{equation} where $k\in (0,1]$ is the proportion of nodes retained, $top[kN]$ is based on the self-attention score $Score$ to select the nodes with the top proportion $k$, and $N$ is the total number of nodes, $top-rank[kN]$ returns the index of the node of $top[kN]$, and then performs an index operation on $Score$ to update the mask $Score_{mask}$.

Next, we perform a pooling operation on feature data by GCN. The new feature matrix and the corresponding adjacency matrix are obtained as follows:

\begin{equation}\label{eq:pabc}
\begin{aligned}
X^{'} = X_{index}\ \ \ \ \ \ \ \ \ \\
X_{out}=X^{'} \odot Score_{mask}\\
A_{out}=A_{index,index}\ \ \
\end{aligned}
\end{equation} where $\odot$ represents the broadcasted element-wise product, $X_{out}$ is the new feature matrix and $A_{out}$ refers to the corresponding adjacency matrix.

Finally, the readout layer is provided to change features to the fixed size before graph classification. The representation results are concatenated by global average pooling and global max pooling, which is shown as follows:
\begin{equation}\label{eq:5}
s=\frac{1}{N} \sum_{i=1}^{N} x_{i} \| \max _{i=1}^{N} x_{i}
\end{equation}
where $N$ is the number of nodes, $X_i$ represents the feature value of $i$-th node, and $\| $ is a concatenation operator.

% \begin{equation}\label{eq:pabc}
% \begin{aligned}
% GCN^l=\sigma (\tilde{L}_{sym}GCN^{l-1}W^{l-1})\\
% GCN^1=\sigma (\tilde{L}_{sym}XW)
% \end{aligned}
% \end{equation}

% \begin{equation}\label{eq:pabc}
% \begin{aligned}
% Score=\sigma (\tilde{L}_{sym}XW_{att})\ \ \ \ \ \ \\
% index=top-rank(Score,[kN])\\
% Score_{mask}=Score_{index}\ \ \ \ \ \
% \end{aligned}
% \end{equation}

\subsection{Domain adversarial based on the graph}
Due to individual differences in emotion, the generalization of the emotion recognition model usually did not perform well. To improve the generalization performance of our model among different subjects, we propose a method of domain adversarial based on the graph.

The main advantage of our method is to reduce the computational complexity significantly. Differing from the work \cite{zhong_eeg-based_2020}, we apply domain adversarial to the graph after self-attention pooling instead to nodes, since this graph after pooling contains the crucial features obtained by extracting from GCN. The input data of the domain adversarial based on the graph are from the readout layer, including the source and target domains.
% In our work, the data from the source domain refers to data from the training set, while the target domain represents the data from the testing set.

The core of domain adversarial based on the graph is the domain classifier. In the training process, the mixed samples from the source domain with labels and the target domain without labels are put into the domain adversarial model. To achieve effective domain migration, the domain classifier in this model is expected not to successfully distinguish between the training (source) domain data and the testing (target) domain data. During training, the domain classifier discriminates between the source and the target domains, and no-crucial features of the source domain and target are removed.

% In other words, in order for the model to be well generalized from one domain to another, it must be ensured that the inside of the neural network does not contain distinctive information about the input source (source domain or target domain). Based on this theory, it is proposed that the domain adversarial training of neural network can be used in any neural network architecture.

% Let $G_f (\cdot ;\theta_f)$ be the D-dimensional neural network feature extractor, the parameter is $\theta_f$, $G_y (\cdot ;\theta_y)$ is the neural network classifier, the parameter is $\theta_y$, The predictive output is $O_y$, $G_d (\cdot ;\theta_d)$ is the neural network domain predictive classifier, the parameter is $\theta_d$, and the predictive output is $O_d$. Then for the input training data $x_i$, $O_y=G_y (G_f (x_i;\theta_f);\theta_y)$, $O_d=G_d (G_f (x_i;\theta_f);\theta_d)$. Suppose the loss function of the neural network classifier is $L_y$ and the domain prediction loss function is $L_d$, then:
% % L_d \left( G_d (G_f (x_i;\theta_f);\theta_d), d_i \right)

% \begin{equation}\label{eq:pabc}
% \begin{aligned}
% L_y^i(\theta_f, \theta_y)=L_y (O_y,y_i)\\
% L_d^i(\theta_f, \theta_d)=L_d (O_d, d_i)
% \end{aligned}
% \end{equation}

For the loss function that needs to be optimized, we select $X^S$, $X^T \subseteq GCN_{ferture}\in \mathbb{R}^{G\times d}$ from $GCN_{ferture}$, where $X^S$ is the source domain data, $X^T$ is the target domain data, $G$ is the number of graphs, $d$ is the dimension of the graph data. The data label belonging to the source domain is set to 0, and the data label belonging to the target domain is set to 1. It is converted to one-hot form as $Y^S_i =[1,0]$, $Y^T_i =[1,0]$, and then cross entropy (CE) is employed to construct as follows:

\begin{equation}\label{eq:pabc}
\begin{split}
E_D & = H(Y^S,q^S)+H(Y^T,q^T)\\
    & =-\left( \sum_{i=1}^{G^T} \sum_{j=1}^C Y^S_{i}(x_j)log q_i(x_j) + \right.\\
    & \quad \left. \sum_{i=1}^{G^S} \sum_{j=1}^C Y^T_{i}(x_j)log(q_i(x_j) \right)
\end{split}
\end{equation}

Classifier in domain adversarial based on the graph is a three-layer fully connected neural network for emotion recognition, which attempts to find the correct emotion based on features from graph self-attention pooling. Kullback-Leibler (KL) divergence is adopted as the loss function of the emotion recognition classifier, since it can measure how one probability distribution is different from another.

\begin{equation}
\centering
\begin{split}
D_{K L}(\hat{Y} \| q) = \sum_{i=1}^{N}\left[\hat{Y}(x_{i}) \log \hat{Y}(x_{i})- \right.\\
\left. \hat{Y}(x_{i}) \log q\left(x_{i}\right)\right]
\end{split}
\end{equation}
where $\hat{Y}$ represents the real data, a measured probability distribution. Distribution $ q(x_{i})$ represents instead a theory distribution of the data.

The training process is to minimize $E_{all}$, which is sum of the emotion recognition loss ($L_y^i$) and the domain classification loss ($L_d^i$).
\begin{equation}\label{eq:pabc}
% \centering
\begin{aligned}
L_y^i(\theta_f, \theta_y)=D_{KL}(\hat{Y}||q) \\
L_d^i(\theta_f, \theta_d)=E_D=H(Y^S,q^S)+H(Y^T, q^T)\\
E_{all}=L_y^i(\theta_f, \theta_y)+L_d^i(\theta_f, \theta_d)\\
\end{aligned}
\end{equation}

% Due to the differences between individuals, the generalization of the emotion recognition classification model is usually poor. To improve the generalization performance of our model among different subjects, we propose a method of graph domain confrontation. Compared with adversarial training, to minimize the distribution difference between the source domain (training set) and target domain (test set) data. More specifically, we will extract the features of the data and the GCNf erture representation score obtained by coarsening the graph. For the source domain and the target domain, in the training process, the source domain and the target domain are confused by the domain confrontation model, and the features in the source domain and the target domain that are irrelevant to the features required by the model are removed. For the loss function that needs to be optimized, we Select XS from GCNf erture, XT⊆GCNf erture∈RG×d, where XS is the source domain data, XT is the target domain data, G is the number of graphs, d is the dimension of the graph data, and will belong to the source domain data The label is set to 0, the data label belonging to the target domain is set to 1, and the conversion to one-hot form is YSi= [1,0], YTi= [1,0], and then the cross entropy is used for the following construction

\section{Experiments and Evaluation}
To evaluate our model, we apply DAGAM to two public emotion EEG-based datasets: SJTU Emotion EEG Dataset (SEED) \cite{zheng_investigating_2015}, and an evolution of the original SEED dataset (SEED IV) \cite{zheng_emotionmeter_2019}.

\subsection{Implementation details}

In the experiment for two datasets, we set hyper-parameters in DAGAM as follows: the number of GCN layers $L$ is 3; the pooling ratio $k$ used in self-attention pooling is set to 0.5. The classifier of emotion recognition based on the graph is a three-layer fully connected neural network. The Adam is used as the model's gradient descent optimizer with the value of $0.001$. We implemented the whole model by PyTorch. The model runs on the server with Intel Core i9-9900K CPU @ 3.60GHz, 32GB memory, 512GB SSD, and NVIDIA GeForce RTX 3090 running Linux Ubuntu 18.04.03LTS.

\subsection{Dataset instruction}
These datasets collect EEG signals by the same device: ESI NeuroScan with 62 channel electrodes according to the international 10-20 system at a sampling rate of 1000Hz. The raw EEG signals from these datasets are preprocessed and extract different salient features based on previous studies \cite{shi_differential_2013}. The detailed information is provided as follows.

\subsubsection{SEED and SEED IV}
% \subscetion{Data Prepossessing}
In the SEED, 15 film clips were chosen to invoke three kinds of emotions: positive, neutral, and negative. Fifteen subjects participated in the experiment. There were 15 trials for each subject in the experiment. In the SEED IV, 72 film clips were chosen to invoke four kinds of emotions: happy, sad, fear, or neutral. It had also recruited 15 subjects to participate in this experiment. Three sessions, including 24 trials, were performed on different days for each subject. The raw EEG data were downsampled at 200Hz to facilitate recognition. Then a bandpass filter with 1Hz to 75Hz was applied to remove the noise and artifacts. In our experiments, a time-frequency domain feature, called differential entropy (DE) \cite{shi_differential_2013} was extracted.
\begin{equation}\label{eq:pabc}
\begin{split}
h(X) & =-\int_{\infty}^{\infty} \frac{1}{\sqrt{2 \pi \sigma^{2}}} \ell^{-\frac{(x-\mu)^{2}}{2 \sigma^{2}}} \log \left(\frac{1}{\sqrt{2 \pi \sigma^{2}}} \ell^{-\frac{(x-\mu)^{2}}{2 \sigma^{2}}}\right)dx \\
     & =\frac{1}{2} \log \left(2 \pi \ell \sigma^{2}\right)
\end{split}
\end{equation}
where the time series $X$ follows the Gauss distribution $N(\mu, \sigma^{2})$.

\subsection{Performance analysis}
We compare DAGAM with other baseline methods to comprehensively evaluate our model, including the state-of-the-art (SOTA) in mean accuracy and standard deviation for SEED and SEED IV, respectively. The confusion matrix analysis is provided following. It ends with the ablation study.

\subsubsection{Subject-independent emotion recognition}
We conduct experiments on two datasets (SEED and SEED IV) using Leave-one-out cross-validation (LOOCV) to evaluate the performance of DAGAM on subject-independent emotion recognition. The experiment settings are followed by \cite{li_bi-hemisphere_2021,zhong_eeg-based_2020}, which tests our DAGAM on one subject and trains on the remaining subjects for each fold. LOOCV evaluates every subject in datasets. The mean accuracy (ACC) and standard deviation (STD) are compared.

The performance of our DAGAM is shown in Table \ref{tb:comparsion in SEED and IV}, which lists the comparison between the DAGAM model and other methods in the subject-independent in SEED and SEED IV. The comparison includes 17 methods as follows: KLIEP \cite{kanamoriLeastsquaresApproachDirect2009}, ULSIF \cite{kanamoriLeastsquaresApproachDirect2009}, STM \cite{chuSelectiveTransferMachine2017}, SVM \cite{suykensLeastSquaresSupport1999},TCA \cite{panDomainAdaptationTransfer2011}, SA \cite{fernandoUnsupervisedVisualDomain2013},GFK \cite{gongGeodesicFlowKernel2012}, A-LSTM \cite{song_mped_2019}, T-SVM \cite{collobertLargeScaleTransductive}, DANN \cite{ganin_domain-adversarial_2016}, DAN \cite{liCrossSubjectEmotionRecognition2018}, BiDANN-S \cite{liBihemisphereDomainAdversarial2018},
BiHDM \cite{li_novel_2021}, RGNN \cite{zhong_eeg-based_2020}, WGAN-DA \cite{luoWGANDomainAdaptation2018}, and TDANN \cite{baoTwoLevelDomainAdaptation2021}.

Obviously, our DAGAM performs better than the other 17 methods, including SOTA both on SEED and SEED IV. The DAGAM achieves the highest accuracy with the lowest standard deviation. It improves the accuracy of SOTA by $4.69\%$ for SEED and $6.90\%$ for SEED IV, respectively.

% \begin{table}[!ht]
%     % increase table row spacing, adjust to taste
%     % \renewcommand{\arraystretch}{1.3}
%     % if using array.sty, it might be a good idea to tweak the value of
%     % \extrarowheight as needed to properly center the text within the cells

%     % \label{table_example}
%     \label{tb:comparsion in SEED and IV}
%     \centering
% \begin{tabular}{c| c| c }
%     % \hline
%     \bf{Method} &  \bf{SEED}& \bf{SEED-IV} \\
%     % \hline
%     % \hline
%     KLIEP & 45.17/17.76 & 31.46/9.20 \\
%     ULSIF & 51.18/13.57 & 32.99/11.05 \\
%     STM & 51.23/14.82 & 39.39/12.4 \\
%     SVM & 56.73/16.29 & 37.99/12.52 \\
%     TCA & 63.64/14.88 & 56.56/13.77 \\
%     SA & 69.00/10.89 & 64.44/9.46 \\

% 	GFK & 71.31/14.09 & 64.38/11.41 \\
% 	AS-LSTM & 72.18/10/85 & 55.03/9.28\\
% 	DANN \citep{ganin2016domain} & 75.08/11.18 & 47.59/10.01 \\
% 	DGCNN \citep{song2018eeg} & 79.95/9.02 & 52.82/9.23 \\
% 	DAN \citep{li2018cross}& 83.81/8.56 & 58.87/8.13 \\
% 	BiDANN-S \citep{li2018bi}& 84.14/6.87 & 65.59/10.39 \\
% 	BiHDM \citep{li2020novel}& 85.40/7.53(SOTA) & 69.03/8.66\\
% 	RGCN \citep{zhong2020eeg}& 85.3/6.72 & 73.84/8.02(SOTA) \\
%     \bf{our}& \textbf{  89.92/3.87}&\textbf{ 80.83/5.58}\\
%     % \hline
%     \end{tabular}
%     \caption{Subject-independent classification accuracy (mean/std) on SEED and SEED-IV.ACC/STD(\%)}
% \end{table}

\begin{table}
\centering
\begin{tabular}{lll}
\toprule
\bf{Method} &  \bf{SEED}& \bf{SEED-IV} \\
\midrule
KLIEP & 45.17/17.76  & 31.46/9.20 \\
ULSIF & 51.18/13.57  & 32.99/11.05  \\
STM & 51.23/14.82  & 39.39/12.4  \\
SVM & 56.73/16.29  & 37.99/12.52  \\
TCA & 63.64/14.88  & 56.56/13.77  \\
SA & 69.00/10.89  & 64.44/9.46  \\
GFK & 71.31/14.09  & 64.38/11.41  \\
A-LSTM & 72.18/10/85  & 55.03/9.28  \\
T-SVM & 72.53/14 & --- \\
DANN  & 79.19/13.14 & 47.59//10.01  \\
DGCNN  & 79.95/9.02 & 52.82/9.23  \\
DAN & 83.81/8.56 & 58.87/8.13 \\
BiDANN-S & 84.14/6.87 & 65.59/10.39  \\
BiHDM & 85.40/7.53 & 69.03/8.66 \\
RGNN& 85.3/6.72 & 73.84/8.02(SOTA) \\
WGAN-DA & 87.07/7.14&---\\
TDANN & 87.90/6.13(SOTA)&---\\
\bf{DAGAM}& \textbf{92.59/3.21}&\textbf{80.74/4.14}\\
\bottomrule
\end{tabular}
\caption{Subject-independent emotion recognition results (mean/std) on SEED, and SEED-IV. ACC/STD(\%)}
\label{tb:comparsion in SEED and IV}
\end{table}

We directly quota emotion recognition results of other baselines from the work of \cite{li_bi-hemisphere_2021}. Our model improved substantially the performed much better than others concerning the accuracy, but with a relatively high standard deviation. We have double-checked our results.

In the aforementioned experiments, our DAGAM can further improve the subject-independent emotion recognition compared with other methods. Among the methods compared with our model, there are two methods that use graph neural networks: DGCNN \cite{song_eeg_2020} and RGNN \cite{zhong_eeg-based_2020} and six methods that use domain adversarial training: TDANN \cite{baoTwoLevelDomainAdaptation2021}, WGAN-DA \cite{luoWGANDomainAdaptation2018}, RGNN \cite{zhong_eeg-based_2020}, DAN\cite{liCrossSubjectEmotionRecognition2018},
BiDANN-S\cite{liBihemisphereDomainAdversarial2018}, DANN \cite{ganin_domain-adversarial_2016}.
No one adopted the attention mechanism. Therefore, we assume that the graph self-attention pooling helps effectively to extract crucial invariable features and remove irrelevant ones.

To further verify this assumption, we do further experiments. In each round of the experiments on two datasets, we modify the core hyper parameters in graph attention neural networks: top proportion $k$. Table \ref{tb:comparsion in K} shows the results. It can be easily found that the experimental results have undergone obvious changes, especially on SEED almost $5\%$. So the graph self-attention pooling did play a central role in our model.

% \begin{table}[h!]
% \centering
% \begin{tabular}{lrrr}
% \toprule
% \bf{$k$} &  \bf{SEED}& \bf{SEED-IV} \\
% \midrule
%     0.5& { 89.03/3.84}&{ 79.17/5.56}\\
%     0.3 & 87.26/4.56 & 78.80/5.88 \\
%     0.1 & 87.55/4.57 & 78.81/4.56\\
% \bottomrule
% \end{tabular}
% \caption{Comparison in different top proportion $k$. ACC/STD(\%)}
% \label{tb:comparsion in K}
% \end{table}

\begin{table}[h!]
\centering
\begin{tabular}{lll}
\toprule
\bf{$k$}&\bf{SEED}&\bf{SEED-IV}\\
\midrule
	0.9& {87.70/5.25}&{80.64/3.63}\\
	0.8& {88.59/4.35}&{80.18/4.68}\\
	0.7& {87.56/3.22}&{80.09/4.55}\\
	0.6& {88.15/3.60}&{80.18/4.87}\\
    0.5& \bf{92.59/3.21}&\bf{80.74/4.14}\\
	0.4& {89.57/4.62}&{79.53/4.71}\\
    0.3& 88.57/4.02 & 80.62/6.22\\
	0.2& {87.70/3.88}&{79.72/4.41}\\
    0.1& 88.15/4.19 & 80.55/4.13\\
\bottomrule
\end{tabular}
\caption{Comperision in different top proportion $k$. ACC/STD(\%)}
\label{tb:comparsion in K}
\end{table}

\subsection{Confusion Matrix analysis}
\begin{figure}[!ht]

  \centering
     \begin{subfigure}[b]{0.5\columnwidth}
         \centering
         \includegraphics[width=1\columnwidth]{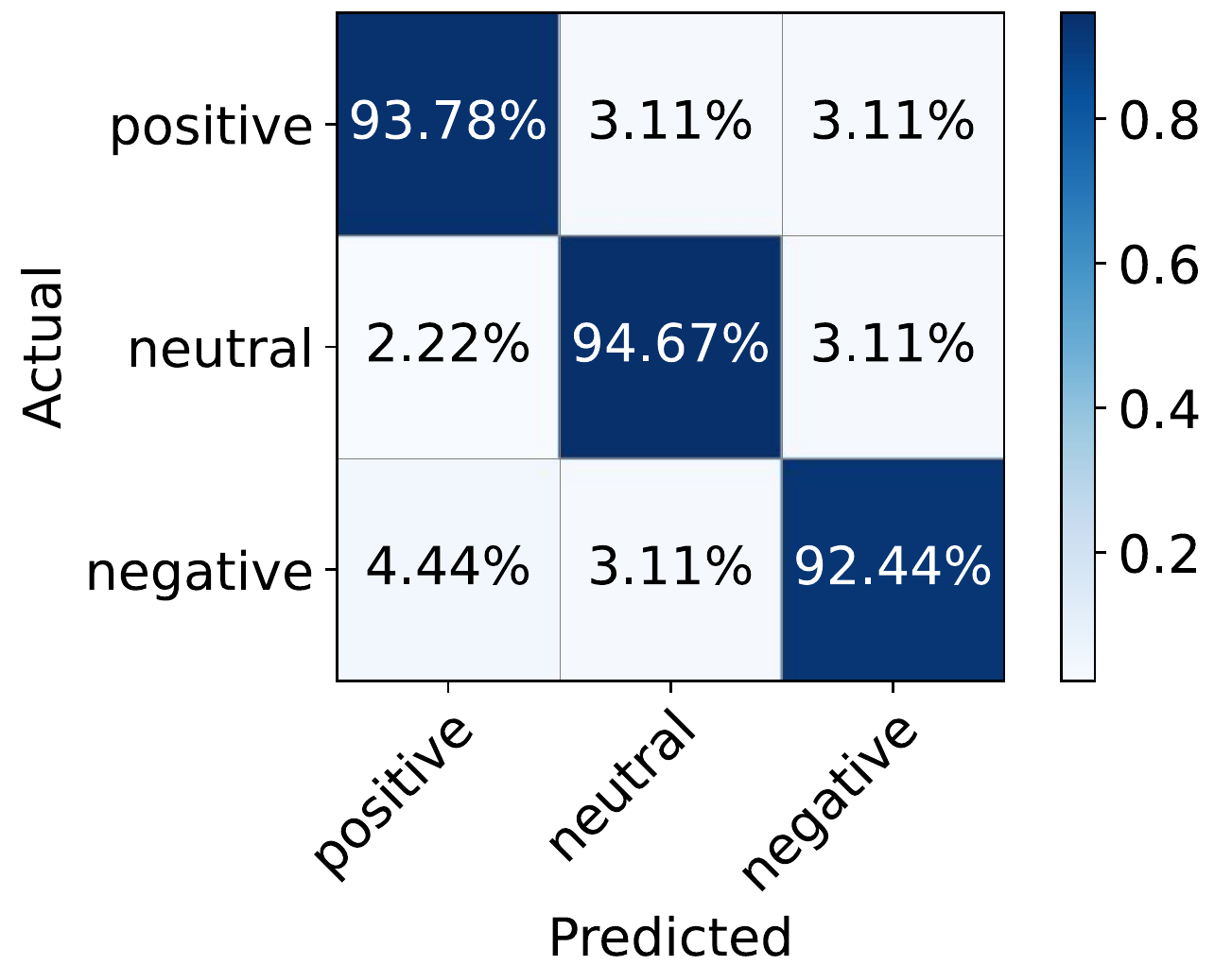}
         \caption{}
         \label{fig:Confusion Matrix (a)}
     \end{subfigure}

     \begin{subfigure}[b]{0.5\columnwidth}
         \centering
         \includegraphics[width=1\columnwidth]{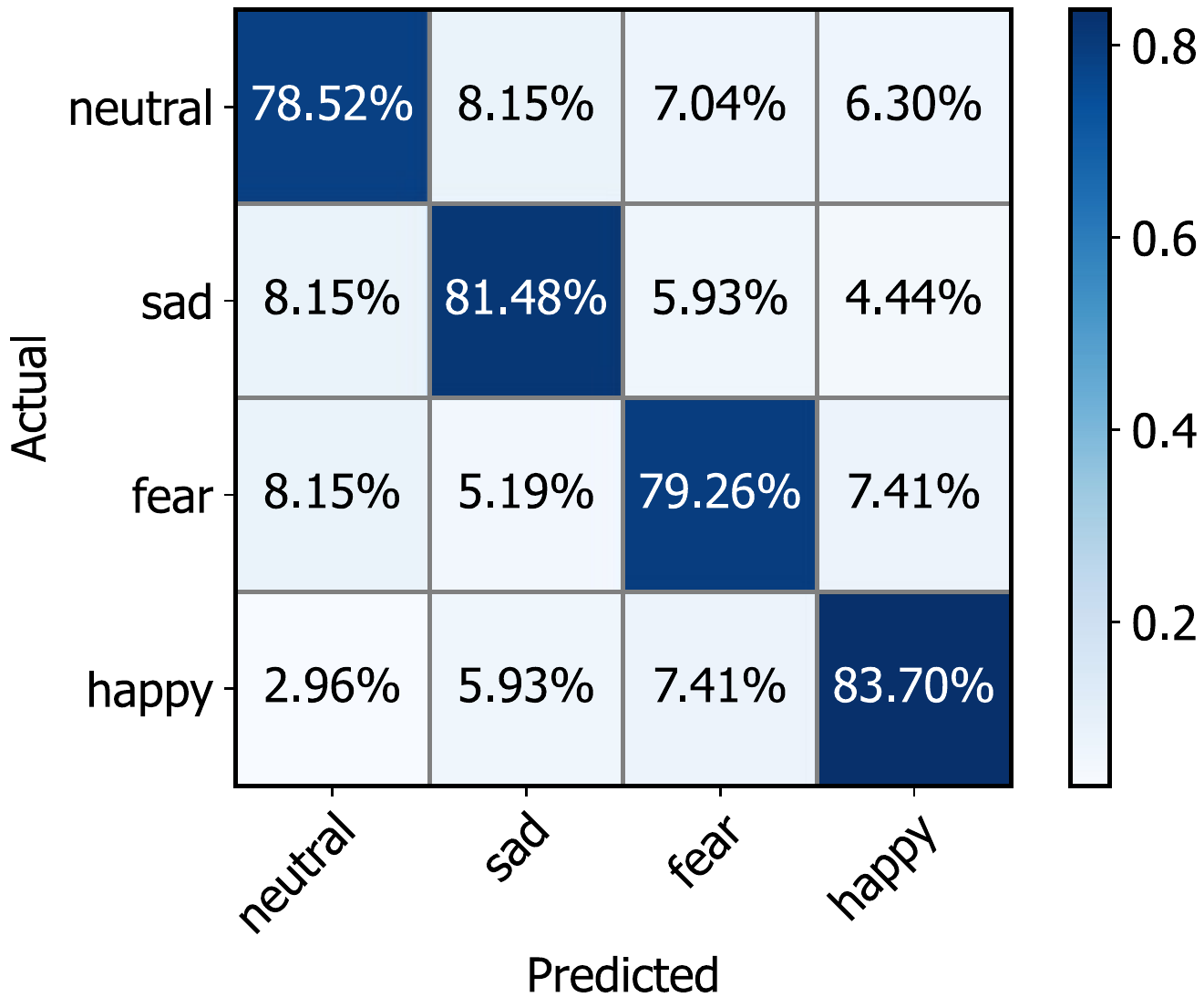}
         \caption{}
         \label{fig:Confusion Matrix (b)}
     \end{subfigure}
    %  \begin{subfigure}[b]{0.7\columnwidth}
    %      \centering
    %      \includegraphics[width=1\textwidth]{mped.pdf}
    %      \caption{}
    %      \label{fig:Confusion Matrix (c)}
    %  \end{subfigure}
   \caption{The confusion matrices of the subject-independent EEG emotion recognition results using our DAGAM on SEED and SEED IV. (a) Confusion matrix of SEED; (b) Confusion matrix of SEED IV.}
     \label{fig:three graphs}
\end{figure}

To make a deep insight into our model for different emotions, we provide the confusion matrix for SEED and SEED IV. As shown in Fig.\ref{fig:three graphs}, these confusion matrices represent in percentage with rows normalized.

For SEED, as shown in Fig.\ref{fig:Confusion Matrix (a)}, our model performs a high-level accuracy for all emotions. It performs on neutral emotions much better than others, while it is not very sensitive to negative emotions. The nearly $10\%$ of neural and negative emotions are misrecognized with negative motion.

For SEED IV, as shown in Fig.\ref{fig:Confusion Matrix (b)}, our model performs around $80\%$ for all four categorized emotions. It is good at distinguishing happy, but weak in recognizing neural. $8.15\%$ of neural emotions are categorized as sad by mistakes, and $7.04\%$ and $6.30\%$ of it are recognized as fear and happy, respectively.

Overall, the model shows a fairly high level of emotion recognition.

\subsection{Ablation study}

% Kullback-Leibler Divergence (KL divergence) that training data and testing data time seems similar but different distributions.

DAGAM mainly adopted three methods in different parts of the model to tackle individual differences. Graph self-attention pooling is adopted in the feature extraction phase to extract crucial features based on biological topology. In the phase of graph classification, KL divergence is adopted to handle inaccurate emotion labels, which can quantify differences between the probability distribution of the training set and the testing set. In the training phase, the domain adversarial based on the graph is an attempt to solve the problem of the same labels with different distributions, that is, domain adaption.

To study the effects of the three core parts in the model, we conducted the ablation study (three further experiments) to verify them. In the first experiment, we disable domain adversarial and only use other parts to recognize emotions to study the effects of domain adversarial based on the graph. In the second experiment, we replace KL divergence with the other loss function: cross-entropy to investigate the effect of KL divergence. In the third experiment, we attempt to find the effects of graph self-attention pooling by disabling domain adversarial and replacing KL divergence.

The results are shown in the Table. \ref{tb:ablation study}. The KL divergence has a significant impact on the performance of the model, especially on SEED. Without the KL divergence, the accuracy of SEED drops by nearly $5.78\%$. The domain adversarial based on the graph affects the performance as well, and the accuracy of the two datasets all decreases. We found that only domain adversarial is applied, and its accuracy does not have a large impact. If only the graph self-attention pooling is applied, it remains a good accuracy on these two datasets. This result verifies our previous assumption again that the graph self-attention pooling is a crucial part of our model.

% \begin{table}
% \centering
% \begin{tabular}{lrrr}
% \toprule
% \bf{Method} &  \bf{SEED}& \bf{SEED-IV} \\
% \midrule
%   \bf DAGAM &\bf{92.59/3.21}&\bf{80.74/4.14}\\
%     - Domain adversarial & 88.29/4.71 & 66.32/3.96 \\
%     - KL divergence & 88.59/3.65 & 76.2/3.7 \\
%     - Domain adversarial\\and KL divergence & 87.7/4.67 & 76.02/6.02 \\
% \bottomrule
% \end{tabular}
% \caption{Ablation study for subject-independent classification accuracy (mean/std) on SEED and SEED-IV, Symbol "-" indicates the following component is removed. ACC/STD(\%)}
% \label{tb:ablation study}
% \end{table}

\begin{table}
\centering
\begin{tabular}{lll}
\toprule
\bf{Method} &  \bf{SEED}& \bf{SEED-IV}\\
\midrule
   \bf DAGAM &\bf{92.59/3.21}&\bf{80.74/4.14}\\
    - Domain adversarial & 87.55/5.18 & 78.33/4.64 \\
    - KL divergence & 86.81/3.48 & 77.50/5.17 \\
    - Domain adversarial\\and KL divergence & 88.00/3.70 & 79.53/4.71\\
\bottomrule
\end{tabular}
\caption{Ablation study for subject-independent classification accuracy (mean/std) on SEED, SEED-IV, Symbol "-" indicates the following component is removed. ACC/STD(\%)}
\label{tb:ablation study}
\end{table}

\section{Conclusion}
This study contributes to the growing area of EEG-based subject-independent emotion recognition by proposing a domain adversarial graph attention model (DAGAM). DAGAM is powerful in learning the relationships among EEG channels based on graph pooling. The use of self-attention pooling benefits extracting salient features for the emotion recognition task. The domain adversarial training based on the graph contributes significantly to tackling the cross-subject EEG variations issue. Extensive experiments on two public datasets (SEED and SEED IV) show that the performance of our model achieves SOTA, providing the highest accuracy and low standard deviation than other competitive baselines. In our future work, we will continue to move along the line of graph models.

\bibliographystyle{named}
\bibliography{ijcai22}

\end{document}